\documentclass[twocolumn,showpacs,superscriptaddress,amsmath,amssymb]{revtex4}
\usepackage{dcolumn}% Align table columns on decimal point
\usepackage{bm}% bold math
\usepackage[dvips]{graphicx}
\usepackage{wrapfig,subfigure}
\usepackage{amssymb}
\usepackage{color}
\usepackage{ulem}

\newcommand{\Q}{{\cal Q}}

\newcommand{\qS}{q_{\cal S}}

\begin{document}

\title{Switching exponent scaling near bifurcation points for non-Gaussian noise}

\author{
Lora Billings}
\affiliation{Department of Mathematical Sciences, Montclair State University, Montclair, NJ 07043}
\author{Ira B. Schwartz}
\affiliation{US Naval Research Laboratory, Code 6792, Nonlinear System Dynamics Section, Plasma Physics Division, Washington, DC 20375}
\author{Marie McCrary}
\affiliation{Department of Mathematical Sciences, Montclair State University, Montclair, NJ 07043}
\author{A. N. Korotkov}
\affiliation{Department of Electrical Engineering, University of California, Riverside, CA 92521}
\author{M. I. Dykman}
\email{dykman@pa.msu.edu}
\affiliation{Department of Physics and Astronomy, Michigan State University, East Lansing, MI 48824}
\date{\today}

\begin{abstract}
We study noise-induced switching of a system close to bifurcation parameter values where the number of stable states changes. For non-Gaussian noise, the switching exponent, which gives the logarithm of the switching rate, displays a non-power-law dependence on the distance to the bifurcation point. This dependence is found for Poisson noise. Even weak additional Gaussian noise dominates switching sufficiently close to the bifurcation point, leading to a crossover in the behavior of the switching exponent.
\end{abstract}


\maketitle

Physical systems display generic features near bifurcation parameter values where the number of stable states changes. In this range
the dynamics is controlled by a slow variable, a soft mode. Its noise-induced fluctuations are comparatively large. They ultimately lead to switching of the system from the stable state. Close to a bifurcation point the switching rate $W$ becomes appreciable even where far from this point it is exceedingly small, for a given noise level (for example, for given temperature). The high sensitivity of the rate to the system parameters has been broadly used to determine parameters of Josephson junctions and Josephson junction based systems \cite{Kurkijarvi1972,Fulton1974,Devoret1987,Han1989}, nanomagnets \cite{Wernsdorfer1997b,Sun2001,Krivorotov2004}, mechanical nanoresonators \cite{Aldridge2005}, and recently in quantum measurements \cite{Siddiqi2004,Lupascu2007,Metcalfe2007}.

The analysis of switching conventionally relies on the assumption that the underlying noise is Gaussian. Then the switching exponent ${\cal Q}$, i.e., the exponent in the expression for the switching rate $W\propto \exp(-{\cal Q})$, displays a power-law dependence on the distance to the bifurcation point in the parameter space $\eta$, ${\cal Q}\propto \eta^{\xi}$ \cite{Neel1955,Kurkijarvi1972,Victora1989}. For systems in thermal equilibrium $\xi=3/2$ for a saddle-node bifurcation and $\xi=2$ for a pitchfork bifurcation. This applies also to systems far from equilibrium \cite{Dykman1980,Tretiakov2003}.

Recently, there has been much interest in large fluctuations and switching induced by non-Gaussian noise \cite{Mckane1989,Fuentes2002,Pilgram2003,Billings2008}. Such switching can be used to determine the noise statistics \cite{Tobiska2004,Pekola2004,Ankerhold2007a,Sukhorukov2007,Timofeev2007,Grabert2008,LeMasne2009}. However, the features of the switching rate near bifurcation points have not been explored. Yet, one may expect that the $\eta$-dependence of the switching exponent will differ from that for a Gaussian noise and will be very sensitive to the noise statistics.

In this paper we study the behavior of the switching exponent ${\cal Q}$ for systems driven by Poisson noise. Such noise is often encountered in photon statistics and in the statistics of current in mesoscopic conductors. We show that the scaling is not described by a simple power law, and the overall $\eta$ dependence of ${\cal Q}$ is much weaker than for Gaussian noise. Surprisingly, if in addition to Poisson noise the system is driven even by a comparatively weak Gaussian noise, sufficiently close to the bifurcation point this noise dominates and there occurs a crossover to the standard scaling of ${\cal Q}$ for Gaussian noise.

Generally, one would expect that, unless it is very weak, a Poisson noise would make a stronger effect on the switching rate than a Gaussian noise. This is so, since switching is a rare event on the scale of the characteristic relaxation time of the system $t_r$, it requires a large fluctuation, whose probability is determined by the tail of the noise distribution. Such a tail is less steep for a Poisson noise than for a Gaussian noise.

The ``takeover" of the switching rate by a weak Gaussian noise close to a bifurcation point is a more subtle effect. It emerges because of the qualitatively different ways the fluctuations leading to switching occur for Gaussian and Poisson noises. This can be understood from the equation of motion for the slow variable $q$,
\begin{equation}
\label{eq:Langevin_bif}
\dot q=-U'(q)+f(t).
\end{equation}
Here, $U(q)$ is the effective potential; for the saddle-node and pitchfork bifurcations, $U=U^{\rm (sn)}(q)$ and $U=U^{\rm (pf)}(q)$, respectively \cite{Guckenheimer1987}, with
\begin{equation}
\label{eq:potential}
U^{\rm (sn)}(q)=\eta q-\frac{1}{3}q^3, \qquad U^{\rm (pf)}(q)=-\frac{1}{2}\eta q^2 +\frac{1}{4}q^4.
\end{equation}
For $\eta > 0$ the system has a stable state $q_a$ at the minimum of $U(q)$ (or two such states, for a pitchfork bifurcation) and a saddle point $\qS$ at the local maximum of $U(q)$; for $\eta=0$ these states merge together. The results can be immediately extended also to the case $U^{\rm (pf)}\to - U^{\rm (pf)}$, the subcritical pitchfork bifurcation where for $\eta=0$ a stable state merges with two unstable states.

The force $f(t)$ in Eq.~(\ref{eq:Langevin_bif}) is noise. We will consider the cases where $f(t)$ is a Poisson noise, $f(t)=f_P(t)$ with $f_P(t)=g\sum_n\delta (t-t_n)$, or a white Gaussian noise, $f(t)=f_G(t)$, $\langle f_G(t)f_G(t')\rangle = 2 D \delta(t-t')$, or a combination $f(t)=f_P(t)+f_G(t)$. Both $g$ and $D$ are assumed small, so that the switching rate $W\ll t_r^{-1}=U''(q_a)$. The noises are $\delta$-correlated in time, because the system motion is slow; they are also independent of $q$, since of interest is a small region in the system phase space \cite{Dykman1980}.

We start with the qualitative picture of Poisson-noise induced switching. Here switching occurs only for the appropriate pulse polarity, $(\qS-q_a)/g > 0$. A single noise pulse shifts the coordinate $q$ by pulse area $g$. Switching requires driving $q$ from $q_a$ to $\qS$, from where the system will switch with probability $\sim 1/2$. The necessary $n_0\sim (\qS - q_a)/g$ pulses should occur within time $\lesssim t_r$, so that the system cannot relax back to the attractor between the pulses. The probability of such pulse sequence is equal to $(\nu t_r)^{n_0}\exp(-\nu t_r)/n_0!$, where $\nu$ is the average pulse frequency. By construction, this is the probability to switch in time $t_r$, it is $\sim Wt_r$, and thus gives the switching rate. The corresponding estimate of the switching exponent ${\cal Q}={\cal Q}_P$ for $n_0\gg 1,\nu t_r$ is
\begin{equation}
\label{eq:Q_estimate_Poisson}
{\cal Q}_P\sim \left[(\qS-q_a)/g\right]\ln\left[(\qS-q_a)/g\nu t_r\right].
\end{equation}
From Eq.~(\ref{eq:potential}), $|\qS-q_a|\propto \eta^{1/2}$, whereas $t_r\sim \eta^{1-\xi}$ with $\xi =3/2$ and $\xi=2$ for the saddle-node and pitchfork bifurcations. It is seen from Eq.~(\ref{eq:Q_estimate_Poisson}) that ${\cal Q}_P\propto \eta^{1/2}$ but contains a large $\eta$-dependent logarithmic factor; the ratio $(\qS-q_a)/g\nu t_r$ is the large parameter of the theory.

White Gaussian noise $f_G(t)$, on the other hand, leads to switching by providing a force that overcomes the deterministic force $U'(q)$. From Eq.~(\ref{eq:potential}), $|U'(q)|\lesssim \eta^{\xi-1/2}$ for $q_a<q<\qS$. Since the probability of noise realization is $\propto \exp\left[-\int dt f_G^2(t)/4D\right]$ \cite{FeynmanQM} and the duration of the needed noise outburst is $\sim t_r$, by setting this probability to be $\sim Wt_r$ and $f_G\sim U'$, we obtain for the switching exponent ${\cal Q}={\cal Q}_G$ the familiar expression \cite{Victora1989,Dykman1980}
\begin{equation}
\label{eq:Q_estimate_Gauss}
{\cal Q}_G=C\eta^{\xi}/D
\end{equation}
($C=4/3$ and $C=1/4$ for the saddle-node and pitchfork bifurcation, respectively, see below).

With decreasing $\eta$, the distance between the stationary states $\qS -q_a \propto \eta^{1/2}$ decreases slower than the deterministic force, $|U'(q)|\lesssim \eta^{\xi-1/2}$. Therefore an outburst of Poisson noise required for a transition decreases slower than that of Gaussian noise. Respectively, as seen from Eqs.~(\ref{eq:Q_estimate_Poisson}) and (\ref{eq:Q_estimate_Gauss}), the switching exponent decreases much slower for Poisson noise than for Gaussian noise. As a result, for small Gaussian-noise intensity $D\ll  g$, the switching rate is determined by Gaussian noise for sufficiently small $\eta$, whereas for larger $\eta$ it is determined by Poisson noise.

\begin{figure}[h]
\begin{center}
\includegraphics[width=3.0in]{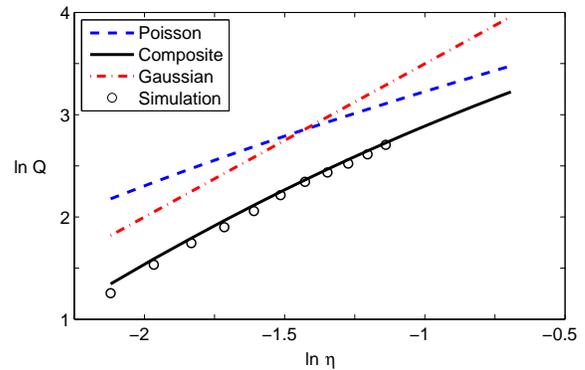}
\end{center}
\caption{Color online. The crossover of the switching exponent ${\cal Q}$ from the Poisson-noise to Gaussian-noise dominated behavior with the decreasing distance $\eta$ to the saddle-node bifurcation point. The solid line shows the result of Eqs.~(\ref{eq:hamiltonian}) and (\ref{eq:barrier_general}) for the Gaussian noise intensity $D=0.009$, the average Poisson pulse frequency $\nu=0.09$, and the Poisson pulse amplitude $g=0.185$; the data points are the results of numerical simulations in which ${\cal Q}$ was found from $W$ using the Gaussian-noise prefactor, $W\approx (\eta^{1/2}/\pi)\exp(-{\cal Q})$. The dashed and dotted lines were obtained from Eqs.~(\ref{eq:hamiltonian}) and (\ref{eq:barrier_general}) by setting $D=0$ or $g=0$, respectively.}
\label{fig:crossover}
\end{figure}

The crossover between Gaussian- and Poisson-noise dominated switching is clearly seen in Fig.~\ref{fig:crossover}. It presents the results on switching near a saddle-node bifurcation point in the presence of both noises. Close to the bifurcation point the slope of $\ln\Q$ vs. $\ln\eta$ is 3/2, as for Gaussian noise. However, as $\eta$ increases the slope approaches that for purely Poisson-noise.

We now give a quantitative theory of the switching exponent and find its $\eta$-dependence for Poisson noise as well as study the crossover between the Poisson- and Gaussian-noise asymptotic. As a starting point, we use a generalized Fokker-Planck equation for the probability density of the system, which follows from Eq.~(\ref{eq:Langevin_bif})
\begin{eqnarray}
\label{eq:Fokker_Planck}
\partial_t\rho =
\partial_q\left[U'(q)\rho(q)\right]+ D\partial^2_q \rho(q) + \nu \left[\rho (q-g) - \rho (q)\right].
\end{eqnarray}
The last two terms in this equation describe the effect of the Poisson noise, i.e., of uncorrelated pulses with average frequency $\nu$ that shift the coordinate by $g$. These terms have the same form as reaction terms in the master equation for a reaction system, with $q$ and $g$ being the number of species and the change of this number in a reaction, respectively \cite{vanKampen_book}, except that in the present case $q$ is continuous; the analogy applies where $g$ is small compared to the typical scale of $q$, in particular compared to $\qS - q_a$. In this case a Poisson noise in the Langevin equation mimics reactions in reaction systems.

For small $D$ and $g$, the switching rate is determined by the probability current away from the initially occupied attraction basin \cite{Kramers1940}. The current is independent of time for $t_r\ll t \ll W^{-1}$ and, to logarithmic accuracy, is given by the quasistationary probability distribution at the saddle point. To find this distribution, we solve Eq.~(\ref{eq:Fokker_Planck}) in the eikonal approximation: we set $\rho(q)=\exp[-s(q)]$, assume that $s(q)-s(q_a)\gg 1$, and keep the leading order terms in $s$ [for example, we disregard $\partial^2_qs$ compared to $(\partial_qs)^2$]. This leads to the equation for $s(q)$ of the form $H(q,\partial_qs)=0$, where
\begin{eqnarray}
\label{eq:hamiltonian}
H(q,p) = - p U'(q) + D p^2 - \nu \left[1- \exp(gp)\right].
\end{eqnarray}
Equation (\ref{eq:hamiltonian}) maps the problem of the quasistationary distribution of the fluctuating system onto the problem of Hamiltonian dynamics of an auxiliary conservative system with coordinate $q$, momentum $p$, and Hamiltonian $H(q,p)$. The switching exponent is ${\cal Q}=s(\qS)-s(q_a)$, or
\begin{eqnarray}
\label{eq:barrier_general}
{\cal Q}= \int\nolimits_{\tilde q_a}^{\tilde q_{\cal S}} p(q)dq.
\end{eqnarray}
Here, $\tilde q_a, \tilde q_{\cal S}$ are the  stationary states shifted to allow for the nonzero mean of the Poisson noise, $\tilde q_i- q_i\approx \nu g/U''(q_i)$ with $i=a,{\cal S}$. The momentum $p(q)$ in Eq.~(\ref{eq:barrier_general}) is the nontrivial solution of equation $H(q,p)=0$, it gives the trajectory of the auxiliary system which goes from $\tilde q_a$ to $\tilde q_{\cal S}$. Equations (\ref{eq:hamiltonian}), (\ref{eq:barrier_general}) could be obtained also by finding the probability density of the most probable realization of noise $f(t)$ necessary to drive the system from $q_a$ to $\qS$ \cite{Billings2008,Sukhorukov2007,Dykman1990}, albeit such calculation would be somewhat more involved.

For purely Poisson noise, i.e., where $D=0$, for small pulse area $g$, equation $H(q,p)=0$ gives $p\approx g^{-1}\{\ln v(q) + \ln[\ln v(q)]\}$, where $v(q)=U'(q)/\nu g$. This estimate applies provided $v(q)\gg 1$; we disregarded higher-order corrections $\propto 1/\ln[v(q)]$. The condition $v(q)\gg 1$ holds in much of the region between $q_a$ and $\qS$ except the immediate vicinities of $q_a,\qS$, i.e., for $|q-q_a|, |q-\qS| \gg \nu |g|/U''(q_a)$. This follows from the estimate $|U'(q)|\lesssim \eta^{\xi-1/2}$ in the central part of the interval $(q_a,\qS)$ and the inequality $(\qS-q_a)/g  \gg \nu t_r=\nu/U''(q_a)$ discussed above.

\begin{figure}
\includegraphics[width=3.0in]{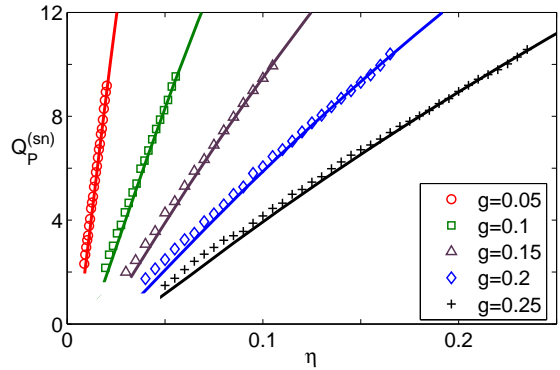}
\caption{Color online. The switching exponent ${\cal Q}_P^{\rm (sn)}$ for a Poisson noise as a function of the distance $\eta$ to the bifurcation point for the saddle-node bifurcation. The mean frequency of noise pulses is $\nu=0.1$. The solid lines are obtained from Eq.~(\ref{eq:barrier_general}) using a numerical solution of equation $H(q,p)=0$. The data points show the results of numerical simulations of switching; the plotted quantity is $\ln(\eta^{1/2}/W\pi)$.}
\label{fig:saddle_node_sim}
\end{figure}

\begin{figure}
\includegraphics[width=3.0in]{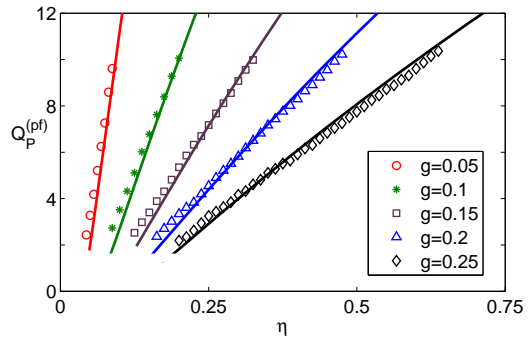}
\caption{Color online. The switching exponent ${\cal Q}_P^{\rm (pf)}$ for a Poisson noise as a function of the distance $\eta$ to the bifurcation point for the pitchfork bifurcation. The mean frequency of noise pulses is $\nu=0.05$. The solid lines are obtained from a numerical solution of equation $H(q,p)=0$ followed by integration, Eq.~(\ref{eq:barrier_general}). The data points show the results of numerical simulations of switching; the plotted quantity is $\ln(\eta/W\pi\sqrt{2})$.}
\label{fig:pitchfork_sim}
\end{figure}

Keeping in $p(q)$ the leading order term and replacing $U'(q)$ by its maximal value between $q_a$ and $\qS$, from Eq.~(\ref{eq:barrier_general}) we obtain the following estimates for the switching exponent for the saddle-node and pitchfork bifurcation, respectively,
\begin{eqnarray}
\label{eq:barriers_Poisson}
&&{\cal Q}_P^{(\rm sn)} \approx \left(2\eta^{1/2}/g\right) \ln \left(\kappa^{\rm (sn)}\eta/\nu g\right), \nonumber\\
&&{\cal Q}_P^{(\rm pf)} \approx \left(\eta^{1/2}/g \right)
\ln \left( \kappa^{\rm (pf)}\eta^{3/2}/\nu\,g \right).
\end{eqnarray}
The parameters $\kappa^{\rm (sn)}, \kappa^{\rm (pf)}$ in the arguments of the logarithms are $\sim 1$. A simple choice $\kappa^{\rm (sn)}=2$ and $\kappa^{\rm (pf)}=1$ gives a close agreement of Eq.~(\ref{eq:barriers_Poisson}) with the results obtained by numerically solving equation $H=0$ and Eq.~(\ref{eq:barrier_general}), which are shown in Figs.~\ref{fig:saddle_node_sim} and \ref{fig:pitchfork_sim}. For the chosen $\nu$, the difference is $<10$\% for $\Q\gtrsim 20$ and $g\geq 0.1$ . Equations (\ref{eq:barriers_Poisson}) justify the estimate Eq.~(\ref{eq:Q_estimate_Poisson}).

The power-law factor $\eta^{1/2}$ in Eqs.~(\ref{eq:barriers_Poisson}) is the same for both types of the bifurcation points \cite{Zou2010}. It is determined simply by the distance between $\qS$ and $q_a$. However, the arguments of the logarithms are different. The logarithmic factors significantly change the switching exponent compared to a simple power-law scaling ${\cal Q}_P\propto \eta^{1/2}/g$. For example, for the pitchfork bifurcation for $g=0.2$ in Fig.~\ref{fig:pitchfork_sim} the logarithmic factor varies from 3.6 for $\eta = 0.5$ to 2.2 for $\eta=0.2$.

The results of the asymptotic theory were compared with numerical simulations of the system Eq.~(\ref{eq:Langevin_bif}). As seen from Figs.~\ref{fig:saddle_node_sim} and \ref{fig:pitchfork_sim}, they are in excellent agreement. In determining the switching exponent from the switching rate $W$, we used the prefactor in $W$ which coincides with that for white Gaussian noise \cite{Dykman2010}.  We note that, for the supercritical pitchfork bifurcation described by the potential $U^{\rm (pf)}(q)$ in Eq.~(\ref{eq:Langevin_bif}), a unipolar (all pulses of the same sign) Poisson noise leads to switching from only one of the two coexisting stable states at the minima of $U^{\rm (pf)}(q)$. For the subcritical bifurcation described by the potential $-U^{\rm (pf)}(q)$, Poisson noise always leads to decay of a metastable state.

In the opposite case of purely Gaussian noise, $g=0$, the nontrivial solution of equation $H(q,p)=0$ is $p=U'(q)/D$. It leads to a power-law dependence of the switching exponent on $\eta$ described by Eq.~(\ref{eq:Q_estimate_Gauss}) and to the appropriate values of the constant $C$ in this equation.

Equation~(\ref{eq:hamiltonian}) describes quantitatively the crossover from Poisson- to Gaussian-noise dominated switching as $\eta$ approaches the bifurcation value $\eta = 0$. If the Gaussian noise is much weaker than the Poisson noise, $D\ll g$, far from $q_a,\qS$ the momentum $p(q)$ is close to the Poisson-noise ($D=0$) solution provided $\eta$ is not too small, so that $M(\eta)\gg D/g$, with $M(\eta)=\eta^{\xi-1/2}/\ln\left(\eta^{\xi-1/2}/\nu g\right)$ (we use that, near its maximum, $|U'|\sim \eta^{\xi-1/2}$). On the other hand, where $M(\eta)\ll D/g$ the momentum is determined by the Gaussian-noise ($g=0$) solution. The position of the crossover on the $\eta$-axis is given by $M(\eta)\sim D/g$. Not surprisingly, for such $\eta$ the switching exponents for the purely Poisson and purely Gaussian noises become of the same order of magnitude, ${\cal Q}_P\sim {\cal Q}_G$. This argument is confirmed by the data in Fig.~\ref{fig:crossover}.

The situation where a weak Gaussian noise is present even where other sources of noise are dominating is typical for practically any physical system. Such noise very often comes simply from the coupling of the system to a reservoir that leads to energy dissipation. The results of the paper explain why near bifurcation points there is often observed the power-law scaling typical for Gaussian noise even where this noise is comparatively weak.

In conclusion,  we have considered noise-induced switching due to a non-Gaussian noise near two generic types of the bifurcation points: saddle-node and pitchfork. In contrast to  the case of Gaussian noise, where the switching exponent scales as a power of the distance to the bifurcation point $\eta$, for a non-Gaussian noise the exponent generally displays a more complicated dependence on $\eta$. We have found it for a Poisson noise, in which case, along with a power-law factor, the exponent has a large logarithmic factor. It turned out that even a weak additional Gaussian noise becomes the major cause of switching sufficiently close to the bifurcation point. A qualitative and quantitative description of the crossover from Poisson to Gaussian noise controlled switching and of the $\eta$-dependence of the switching exponent are in full agreement with numerical simulations.

LB is supported by ARO grant No.~W911NF-06-1-0320. IBS is supported by the Office of Naval Research. ANK is supported by NSA/IARPA/ARO grant W911NF-08-0336. MID is supported by NSF grant CMMI-0900666.

%\bibliographystyle{apsrev}
%\bibliography{c:/Aaa/Bibtex/md10}

\end{document}